\documentclass{PoS}

\def\inb{nb$^{-1}$}
\def\pt{$p_T$}

\title{Early Searches with Jets with the ATLAS Detector at the LHC}

\ShortTitle{Early Searches with Jets with the ATLAS Detector at the LHC}

\author{\speaker{Georgios Choudalakis}%
        \thanks{On behalf of the ATLAS Collaboration.}\\
       University of Chicago, Enrico Fermi Institute\\
       E-mail: \email{georgios.choudalakis@cern.ch}}

\abstract{We summarize the analysis of high-$p_T$ jets in early $pp$ collisions recorded with the ATLAS detector.  Two searches for new physics are presented: One for dijet resonances, and one for quark contact interactions.  The first search sets the most stringent current limit on the mass of a hypothetical excited quark.}

\FullConference{35th International Conference of High Energy Physics - ICHEP2010,\\
		July 22-28, 2010\\
		Paris France}

\begin{document}

\section{Introduction}

We present two analyses searching for new physics in events with at least two hadronic jets.  One analysis searches for dijet resonances using 296 \inb, and the other for quark contact interactions using 61 \inb\ of ATLAS $pp$ collision data at $\sqrt{s}=7$~TeV.\footnote{Updated versions are available in \cite{ourPRL,ourUpdate} and \cite{chiPLB}.}

The inclusive dijet final state is chosen for its sensitivity to many proposed extensions of the Standard Model, such as  quark compositeness, TeV-scale gravity, and new strong dynamics.  The reported amount of data in the dijet final state are enough to probe a kinematic region never observed previously in collider experiments. 

\section{Search for dijet resonances}

A search for dijet resonances is performed in 296 \inb\ of data.  The signature searched for is a cross-section enhancement in a narrow range of the dijet mass spectrum.  The dijet mass spectrum is the distribution of the invariant mass, $m^{jj} \equiv \sqrt{(E_1+E_2)^2-(\vec{p}_1+\vec{p}_2)^2}$, of the system of two highest-\pt\ jets in each event.


Jets are defined using the anti-$k_T$ clustering algorithm \cite{antiKt} with radius $R=0.6$.  
The input is clusters of calorimeter cells seeded by cells with energy significantly above noise level.
Jet energy is calibrated based on the jet-level calorimeter response estimated by the GEANT simulation of the ATLAS detector \cite{Geant4}.

Events are used from periods with LHC and ATLAS conditions appropriate for calorimetery, tracking, triggering, and luminosity measurement.  Events are accepted through a hardware-level trigger dedicated to single high-\pt\ jets.   At least two jets are required per event, one with $p_T>80$~GeV to be in the fully efficient trigger region, and another with $p_T>30$~GeV, to be in the region of full jet reconstruction efficiency.  The two leading jets are required to satisfy several quality criteria~\cite{jetCleaning} and to lie outside the interval $1.3 < |\eta^{\rm jet} | < 1.8$, where the jet energy is not yet measured in an optimal way.  
Events are required to contain at least one primary vertex within 10~cm from the collision point, defined by at least five tracks.
Events containing a poorly measured jet~\cite{jetCleaning} with $p_T > 15$~GeV are vetoed to prevent potential misidentification of the two leading jets; this affects the event selection by less than 0.5\%.

Finally, the two leading jets are required to have $\left|\eta^{\rm jet}\right| < 2.5$, and to satisfy $\left|\eta^{j_1}-\eta^{j_2}\right| < 1.3$.  
Optimization shows these requirements to suppress high-mass QCD background, thus enhancing the potential signal from sources such as excited quark ($q^*$) decays.


The background is determined by fitting the observed spectrum with the function~\cite{CDF_Mjj}
\begin{equation}
f(x) = p_1 (1 - x)^{p_2} x^{p_3 + p_4\ln x},
\label{eqf}
\end{equation}
where $x\equiv m^{jj}/\sqrt{s}$, and $p_{\{1,2,3,4\}}$ are free parameters.  
This function fits well the $m^{jj}$ in {\sc pythia}, {\sc herwig}, and next-to-leading-order (NLO) perturbative QCD predictions for $p\bar{p}$ collisions at $\sqrt{s} = 1.96$~TeV~\cite{CDF_Mjj}.
The same is shown for in $pp$\ collisions at $\sqrt{s} = 7$~TeV using {\sc pythia} and {\sc alpgen} with the ATLAS {\sc geant4}-based detector simulation.


\begin{figure}
\begin{center}
\includegraphics[width=0.45\textwidth]{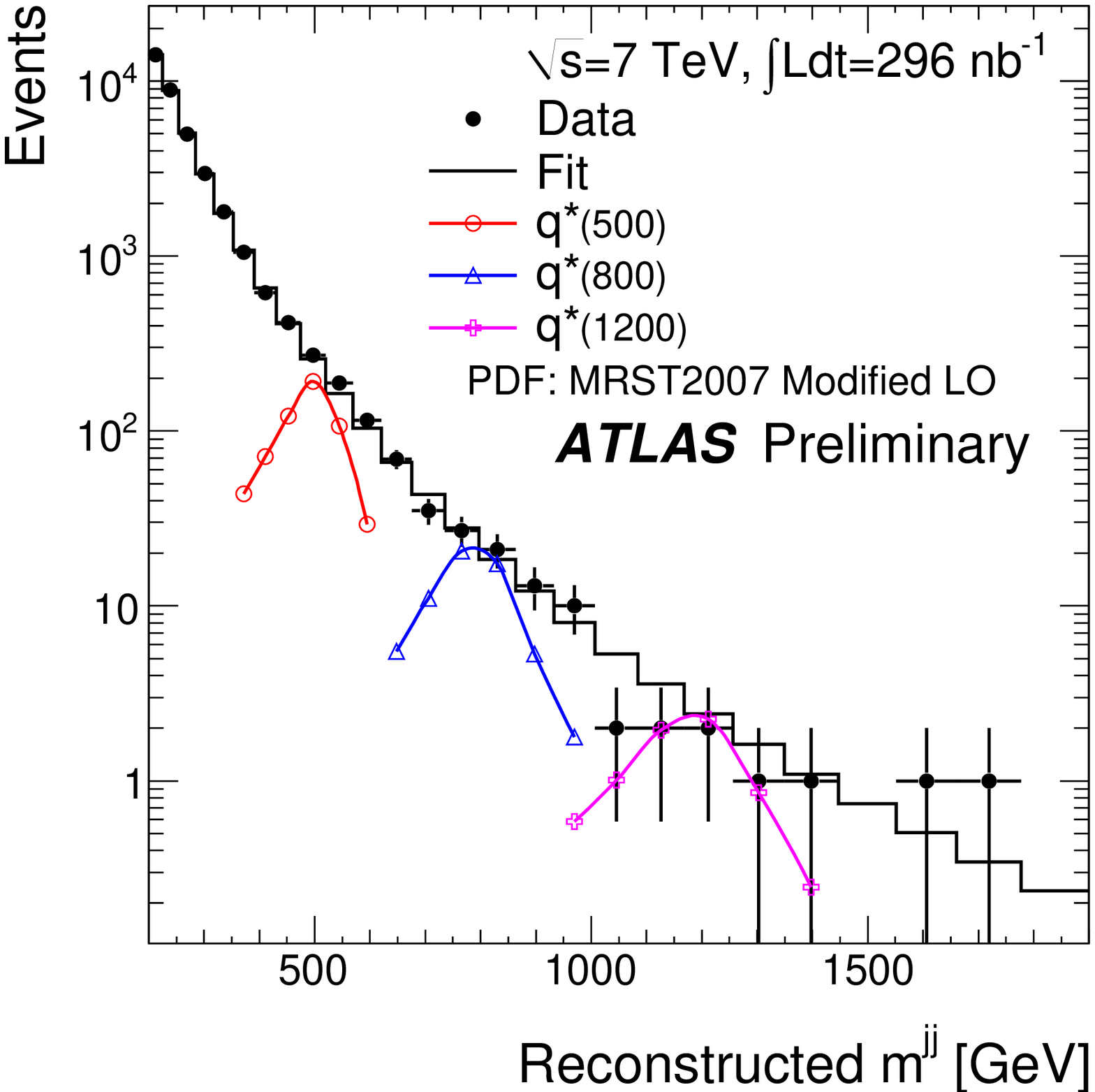}
\includegraphics[width=0.45\textwidth]{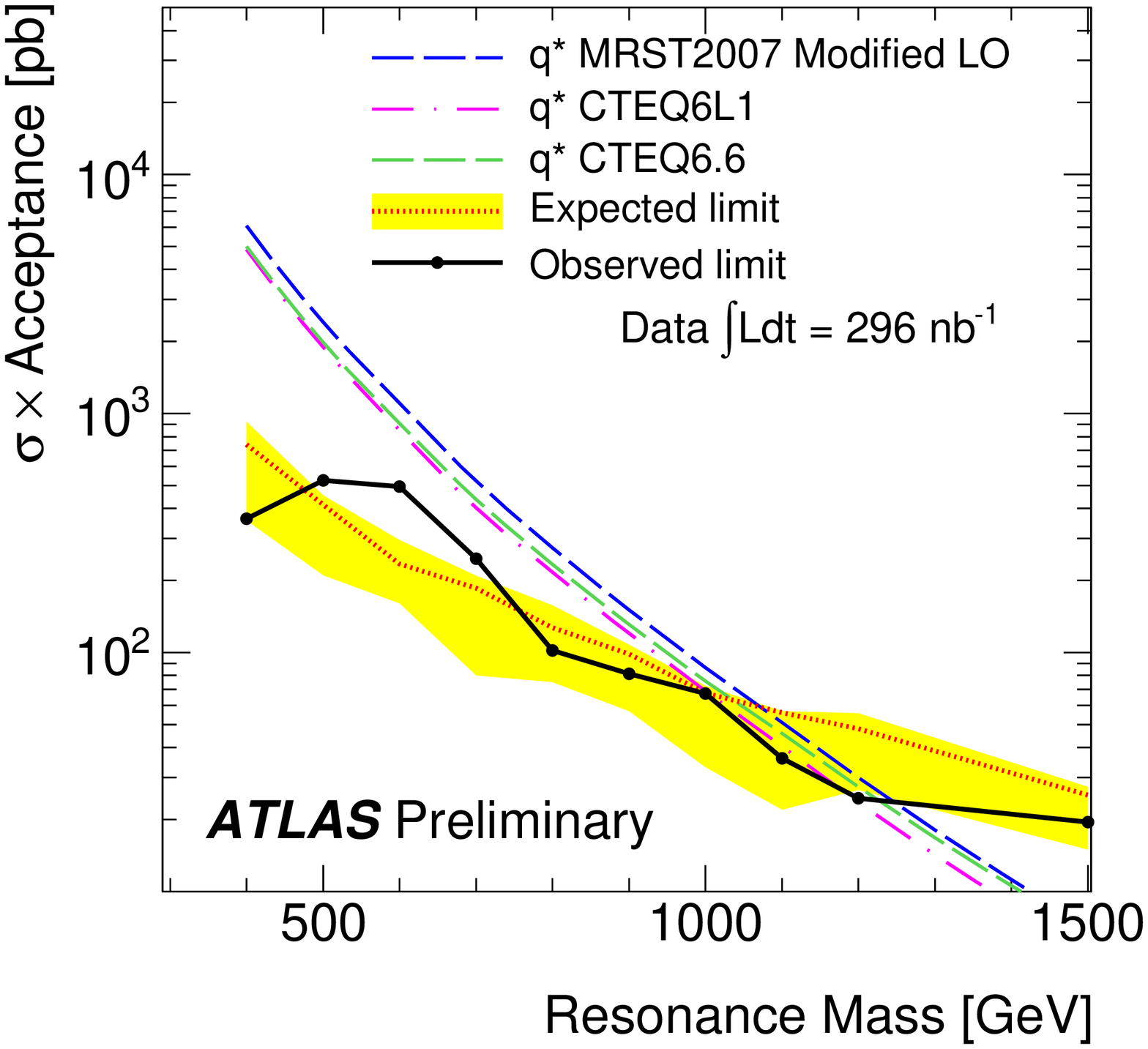}
\caption{{\em Left:} The data dijet mass distribution (filled points) fitted using a binned background distribution described by Eqn. 2.1 (histogram).  The predicted $q^*$ signals for excited-quark masses of 500, 800, and 1200~GeV are overlaid. {\em Right:}  The 95\% CL upper limit on $\sigma\cdot Acceptance$ as a
function of dijet resonance mass (black filled circles).  The red
dotted curve shows the expected 95\% CL upper limit and the
yellow shaded band represents the 68\% credibility interval of the expected limit.  
The dashed curves represent excited-quark $\sigma\cdot Acceptance$ predictions for
different PDF sets. \label{fig:spectrumAndFit}}
\end{center}
\end{figure}

Figure~\ref{fig:spectrumAndFit} (left) shows the observed data after event selection, and the corresponding background.
The level of agreement between data and background is determined using a suite of statistical tests sensitive to resonances, to high-mass excesses of data, and to overall shape discrepancy. 
The results of all tests are consistent with the conclusion that the data are consistent with the fitted parametrization, with $p$-values in excess of 51\%.


A Bayesian approach is used to set 95\% credibility-level (CL) upper limits on $\sigma\cdot Acceptance$ for excited quarks of various test masses.
For each test mass, the background is evaluated from a simultaneous five-parameter fit
of the signal and background distributions to ensure that the
background determination is not biased by the potential presence of the signal considered.  
For every $q^*$\ mass, the posterior probability distribution is computed assuming a uniform prior in the signal yield $s$.  The upper limit for $s$ is then determined by integrating 95\% of the posterior probability distribution.

The dominant sources of systematic uncertainty are the jet energy (6 to 9\% depending on jet $p_T$ and $\eta$), the background fit parameters (reflecting the uncertainty on the parameters resulting from the fit of Eqn.~\ref{eqf} to the data, and ranging from 3\% at low mass to 30\% at high mass), the integrated luminosity (11\%), and the jet energy resolution (assumed 14\%, constant in $p_T$ and $\eta$) \cite{jetProdICHEP}.
These uncertainties are introduced as nuisance parameters into the likelihood function
and marginalized by numerically integrating the product of this modified likelihood, the prior in $s$,
and the priors corresponding to the nuisance parameters to arrive at a modified posterior probability distribution.  

Figure~\ref{fig:spectrumAndFit} (right) depicts the resulting
95\% CL upper limits on $\sigma\cdot Acceptance$ as a function of the
$q^*$ resonance mass after incorporation of systematic uncertainties.
The observed 95\% CL $q^*$ mass lower limit is $1.25$~TeV, using MRST2007 PDFs in the ATLAS default MC09 tune.  
This represents the best limit on the excited quark mass to date, a conclusion which does not depend on varying PDF assumptions, as shown in Figure~\ref{fig:spectrumAndFit} (right).

\section{Search for quark contact interactions}

The search for contact interactions presented here is performed in 61 \inb.
Two observables are used: $\chi$, and $R_\eta$, defined as:
\begin{equation}
\chi \equiv e^{|y_1-y_2|},
\end{equation}
where $y_{1,2}$ is the rapidity of the first two leading jets in \pt, and
\begin{equation}
R_\eta\equiv \frac{N(|\eta_{1,2}|<0.5)}{N(0.5<|\eta_{1,2}|<1.0)},
\end{equation}
where the numerator is the number of events with both leading jets at $|\eta|<0.5$, and the denominator is the number of events with both jets at $\eta$ between 0.5 and 1.

For a contact interaction described by the Lagrangian ${\cal L}_{qqqq}(\Lambda)=\frac{(4\pi)^2}{2\Lambda^2}\bar{\Psi}^L_q\gamma^\mu \Psi_q^L \bar{\Psi}^L_q\gamma_\mu\Psi^L_q$, an excess of data is expected at low $\chi$, in events of high dijet mass.  Similarly, the $R_\eta$ would not be constant versus dijet mass, but would increase at high $m^{jj}$.

The event selection applied is similar to that used in the dijet resonance search, with the following differences:  The leading jet may have $p_T>60$~GeV in the control region (defined below), and the rapidities of the two leading jets have to satisfy $|y_1+y_2|<1.5$, to suppress the effects of PDF uncertainty.

For both $\chi$ and $R_\eta$, the background is obtained from {\sc pythia} QCD after ATLAS detector simulation.  

\begin{figure}
\begin{center}
\includegraphics[width=0.325\textwidth]{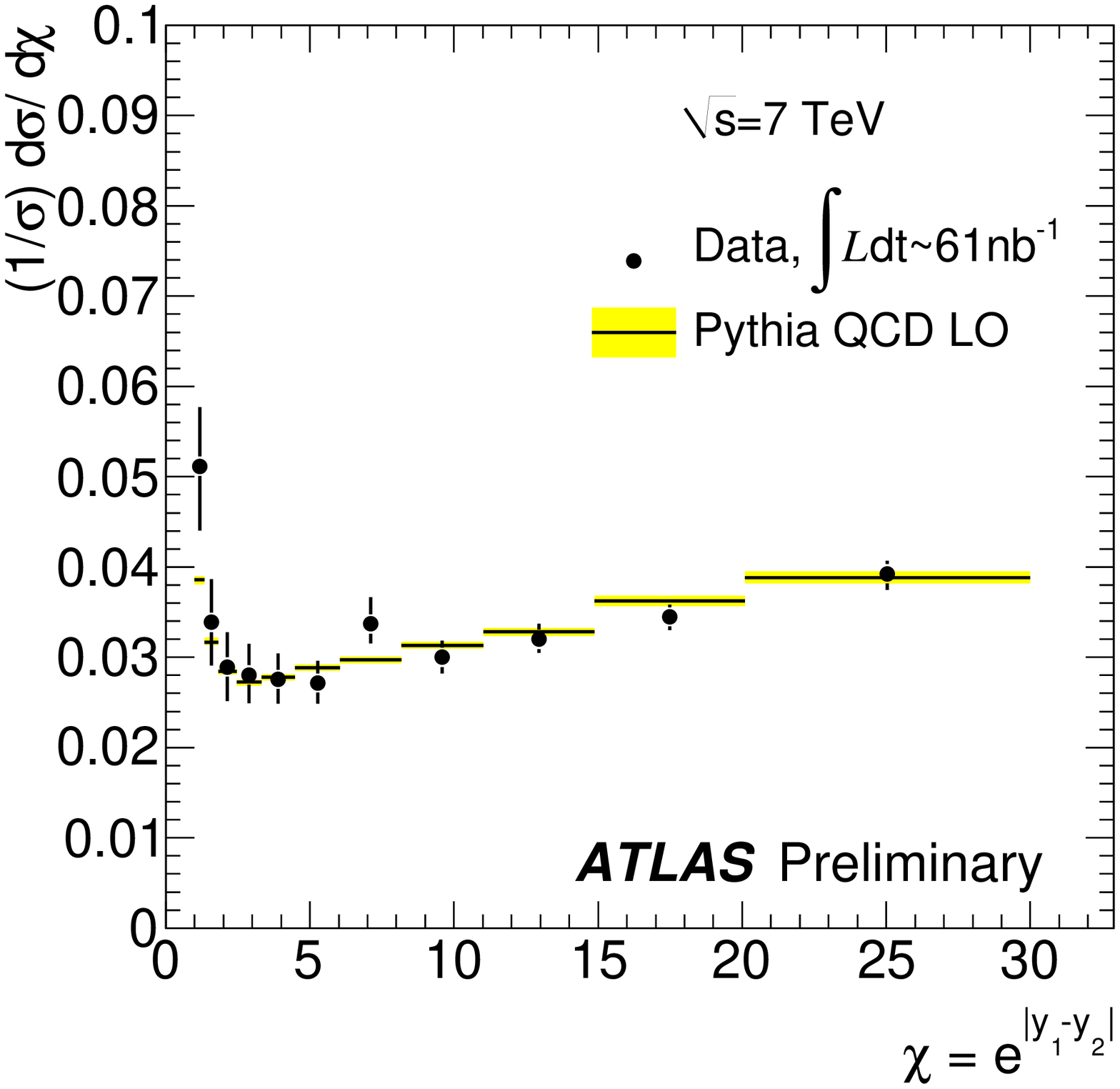}
\includegraphics[width=0.325\textwidth]{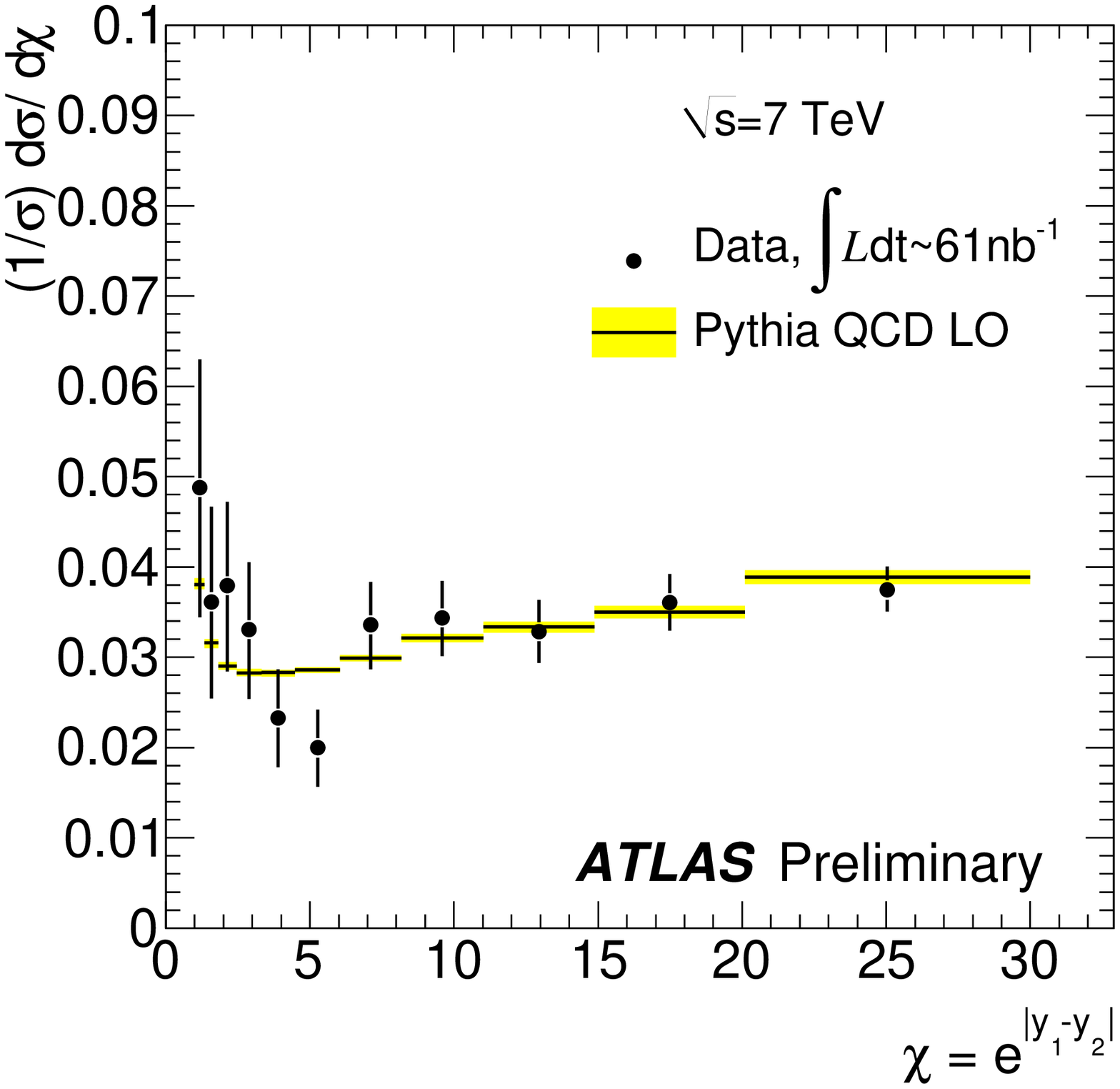}
\includegraphics[width=0.325\textwidth]{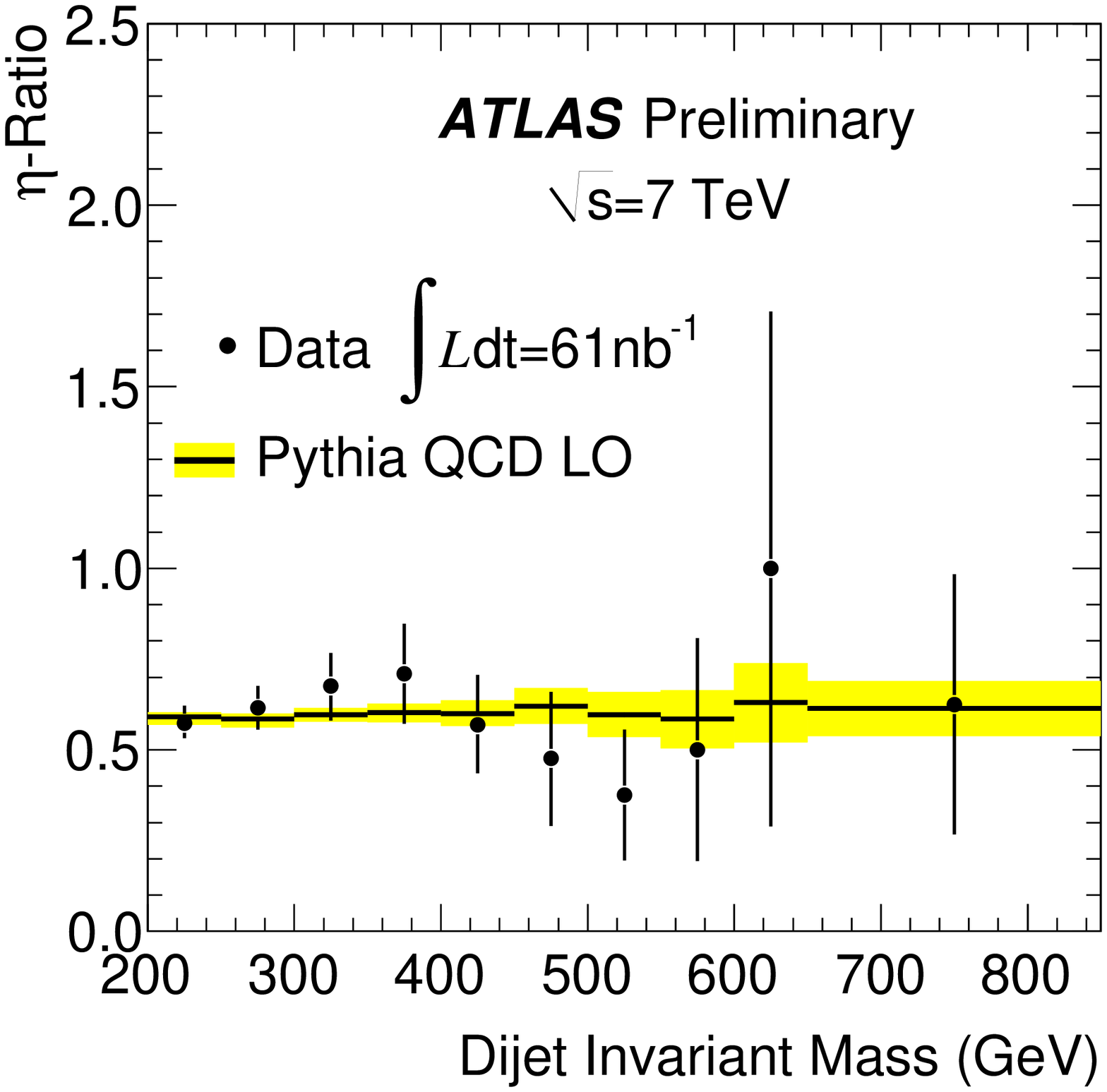}
\caption{{\em Left:} The normalized distribution of $\chi$ in data and LO Standard Model prediction, in the control region where $320 < m^{jj} < 520$~GeV.  {\em Middle:}  Normalized $\chi$ distribution in the signal region, where $520 < m^{jj} < 680$~GeV.  {\em Right:} Average $R_\eta$ (labeled as ``$\eta$-Ratio'') in bins of dijet mass, in data and LO Standard Model prediction. \label{fig:chi}}
\end{center}
\end{figure}

For the analysis of $\chi$, the events are separated in a control region, where $320 < m^{jj} < 520$~GeV, and in a signal region, where $520 < m^{jj} < 680$~GeV.  In the control region good agreement is observed between the data and the background, as shown in Figure~\ref{fig:chi} (left).   In the signal region, shown in Figure~\ref{fig:chi} (middle), the agreement is good, with $\chi^2/NDF=1.07$.

After optimization it was decided to characterize the $\chi$ distribution by the ratio of events in $\chi<4.48$, which is expected to be large for small values of the compositeness scale $\Lambda$.  Using the classical Neyman construction, the observed ratio is translated to a frequentist limit on $\Lambda$, which is: $\Lambda > 930$~GeV at 95\% CL.

A bayesian limit on $\Lambda$ was also computed, computing the likelihood of observing the $\chi$ distribution in Figure~\ref{fig:chi} (middle).  A uniform prior in $1/\Lambda^4$ was used, which corresponds to a uniform probability density in the cross-section of contact interaction.  The bayesian limit obtained is $\Lambda > 875$~GeV, at 95\% CL.

The observed mean $R_\eta$, as a function of $m^{jj}$, shown in Figure~\ref{fig:chi} (right), is consistent with the QCD prediction.  A similar bayesian analysis results in the limit $\Lambda > 760$~GeV at 95\% CL.

\section{Summary}

Two analyses of dijet events are summarized, using the first $pp$ collisions at $\sqrt{s}=7$~TeV recorded with ATLAS.  One analysis is a search for dijet resonances in 296 $nb^{-1}$ of data.  No sign of a resonance is observed in this amount of data.  Limits are set on the mass of a hypothetical excited quark, which serves as a benchmark model for narrow resonances.  The lower limit set on the excited quark mass, assuming MRST2007 PDF set, is 1.25~TeV.  This represents currently the most stringent limit, superseding significantly the latest Tevatron results for the same model \cite{CDF_Mjj}.  It also represents the first new physics result from the LHC to supersede comparable past results.

The second analysis presented is a search for quark contact interactions in 61 \inb\ of data.  The most stringent limit is obtained using the $\chi$ distribution, which results in a frequentist limit on compositeness scale $\Lambda > 930$~GeV at 95\% CL.  
This comparatively low limit is not interpreted as exclusion of a physics model but as a benchmark of the ATLAS sensitivity given the dataset used.

Updated versions of both analyses are available in \cite{ourPRL,ourUpdate} and \cite{chiPLB}.

\end{document}